\documentclass[epj]{svjour}
\usepackage{graphicx,epsf,here,exscale}
\usepackage{amssymb}
\usepackage{amsmath}
\allowdisplaybreaks[2]
\newcommand{\ba}{\begin{array}}
\newcommand{\ea}{\end{array}}
\newcommand{\bc}{\begin{center}}
\newcommand{\ec}{\end{center}}
\newcommand{\be}{\begin{equation}}
\newcommand{\ee}{\end{equation}}
\newcommand{\beq}{\begin{eqnarray}}
\newcommand{\eeq}{\end{eqnarray}}
\newcommand{\bt}{\begin{table}}
\newcommand{\et}{\end{table}}
\newcommand{\btab}{\begin{tabular}}
\newcommand{\etab}{\end{tabular}}

\newcommand{\q}{\quad}
\newcommand{\usp}{{\frac{1}{\pi}}}
\newcommand{\pisd}{{\frac{\pi}{2}}}

\newcommand{\AH}{{\hat A}}
\newcommand{\HH}{{\hat H}}
\newcommand{\NH}{{\hat N}}
\newcommand{\PH}{{\hat P}}
\newcommand{\ZH}{{\hat Z}}

\newcommand{\uni}{{{\hat{\boldmath 1}}}}

\newcommand{\EC}{{\cal E}}
\newcommand{\NC}{{\cal N}}
\newcommand{\HC}{{\cal H}}
\newcommand{\UC}{{\cal I}}

\newcommand{\Nz}{{N_0}}
\newcommand{\PNz}{{\PH_\Nz}}
\newcommand{\eee}[1]{{ {\rm e}^{\normalsize #1}}}

\newcommand{\kphiq}{{\vert\Phi(q)\rangle}}
\newcommand{\bphiq}{{\langle\Phi(q)}}
\newcommand{\kphin}{{\vert\Phi_{\Nz}(q)\rangle}}

\newcommand{\bphind}{{\langle\Phi_{\Nz}(q)}}
\newcommand{\kpsi}{{\vert\Psi_j\rangle}}
\newcommand{\bpsi}{{\langle\Psi_j}}

\newcommand{\DNn}{{\sqrt{\langle\Delta\NH^2\rangle}}}
\newcommand{\DNz}{{\sqrt{\langle\Delta\ZH^2\rangle}}}
\begin{document}
\title{GCM Analysis of the Collective Properties of Lead Isotopes\\
       with Exact Projection on Particle Numbers}
\author{P.-H. Heenen\inst{1} \and 
        A. Valor\inst{1}     \and 
        M. Bender\inst{1}    \and 
        P. Bonche\inst{2}    \and 
        H. Flocard\inst{3}
}
\institute{Service de Physique Nucl{\'e}aire Th{\'e}orique,
           Universit{\'e} Libre de Bruxelles CP229, B--1050 Bruxelles 
           \and
           Service de Physique Th{\'e}orique,
           CE Saclay, F--91191, Giv-sur-Yvette Cedex 
           \and
           Institut de Physique Nucl{\'e}aire, F--91406 Orsay Cedex
}
\authorrunning{P.-H. Heenen \emph{et al.}}
\titlerunning{GCM Analysis of the Collective Properties of Lead Isotopes}
\date{June 1 2001}
\abstract{
We present a microscopic analysis of the collective behaviour of the lead 
isotopes in the vicinity of $^{208}$Pb. In this study, we rely on
a  coherent approach based on the Generator Coordinate Method including 
exact projection on $N$ and $Z$ numbers within a collective space
generated by means of the constrained Hartree-Fock BCS method. With the same 
Hamiltonian used in HF+BCS calculations, we have performed a comprehensive 
study including monopole, quadrupole and octupole excitations as well 
as pairing vibrations. We find that, for the considered nuclei, the 
collective modes which modify the most the conclusions drawn from the 
mean-field theory are the octupole and pairing vibrations.
}
\PACS{{}{}}
\maketitle
%
%
\section{\label{S1}Introduction}
Mean-field methods have been extensively useed to study a wide
range of nuclear properties. There exist now several variants which
although relying on  different physical approaches,
lead to rather similar results.
Among the methods most extensively used in the last years, 
one can mention those based on Skyrme (zero-range) \cite{BFH85,FHS99} 
interactions, or on the Gogny force (finite-range 
interaction) \cite{Gog75,DGo80,PGB00}, the
density-functional approach  \cite{FTT00} 
and the relativistic mean-field method
based on an effective Lagrangian \cite{Rin96,LKR97}.

One of the key ingredient in all these
methods is the way the nucleon-nucleon interaction is treated.
Either a density- functional or a relativistic Lagrangian
is globally adjusted to some classes of nuclei and/or 
nuclear matter properties. Henceforth,  it should be used
without readjustment for any specific problem relevant
to this class of properties. 
Mean-field methods attempt to reproduce 
nuclear many body properties within a description which takes 
into account the one body density only. 
Expectedly, the data accessible to these studies
is mostly limited to that corresponding to one body operators, such 
as the shape deformations or the radii. To that 
already large set one can add
the total binding energy for which the 
Hartree-Fock (HF) or Hartree-Fock Bogoliubov 
(HFB) minimization principle is explicitly designed. 
In practice, other observables are sometimes also included
in the fit of the functionals or lagrangians.  
A typical example is the effective nucleon mass
which is related to the density of single particle 
levels at the Fermi surface.
To obtain a density similar to that deduced
from the results of pickup and 
stripping reactions, the effective mass of the
interaction should be close to the free nucleon mass. On the other
hand, it has been shown that 
RPA correlations which are the simplest type 
of correlations beyond HF, increase
the value of the effective mass \cite{BGi80}.. To account for such
facts, two approaches have been utilized in non-relativistic studies
as soon as the nuclear energy density approach has been shown to
be flexible enough to reproduce nuclear properties with a good
accuracy.
In one of them, one tries to define the
functional so as to avoid a double counting
of the correlations. This implies that the choice of
the effective parameters leaves room for the improvement
expected from a further inclusion of RPA correlations. 
For instance, the mean-field value of the effective
nucleon mass is adjusted to a value smaller than the bare mass
so that RPA correlations may enhance it to the experimental
value. In the past, this has been the most standard approach to
define effective interactions used in the calculation
of the energy functional.
Another strategy consists in the construction 
of functionals which are in a sense ``doubly'' effective
since they are expected 
to take into account the effect of all correlations beyond
mean-field on a specified set of nuclear properties. 
Then, for such properties,
the mean-field calculation should
yield results directly comparable to data.
Two successful examples of the latter strategy are
the interaction of Tondeur et al. \cite{Ton00} 
which reproduces most known nuclear masses with high accuracy
and that of Brown \cite{Bro98} which yields 
single particle level schemes to be used
in subsequent shell model calculations.

In this work, we consider a functional belonging
to the the first family in order to investigate how collective 
modes modify the mean-field properties. 
For this analysis, we rely on the generator 
coordinate method (GCM) taking into account several collective variables.
In the past, such an approach has been shown
to be suited for a description of low lying collective states related to
the quadrupole \cite{KBF96} or the octupole \cite{SHB93} degree of freedom. A
study of the coupling of the dipole and the octupole modes in 
$^{152}$Sm and in the superdeformed band of $^{190}$Hg \cite{HSk96}
has also demonstrated the ability of the GCM to describe in a single
calculation the properties of the ground state, of the low lying collective
states and of the giant dipole resonance. This method seems therefore
to be an appropriate tool to study the effect of correlations beyond
a mean-field approach on the ground state properties of a nucleus.

As  test case, we have selected five Pb 
isotopes around the magic number \mbox{$N=126$}. These nuclei
are in principle well described by mean-field theories. 
The evolution of the two neutron separation energy 
at a magic number is representative
of the shell gap. We will therefore
determine how much it is affected by different types of
collective correlations. Our aim is
to be as systematic as possible within the GCM
and to investigate the effect of all collective
nuclear shape degrees of freedom which are 
believed to be important for these nuclei
: the monopole, quadrupole and octupole modes. 

Moreover, our study
also includes the pairing vibration mode. 
Indeed, as part of this work, we analyze the validity of a
commonly used treatment of pairing correlations beyond 
the BCS theory. 
Since BCS correlations in $^{208}$Pb
vanish, one often attempts
to describe them by means of an approximate 
variation after projection on the nucleon numbers.
In the following, we test the validity of the Lipkin Nogami 
(LN) prescription  \cite{Lip60,NZu64,PNL73}
by performing an exact projection of
the BCS wave functions.
A further study of pairing vibrations also naturally provides
an alternative approximate variation after
projection and allows to test the 
quality of the LN prescription.

The organization of this work is as follows.
In Sect.\ref{S2}, we present the dynamical ingredients 
(Hamiltonian and collective spaces) and 
recall the method while introducing our 
notations. The Sect.\ref{S3} discusses our 
results. In the last section,
we summarize the major conclusions and indicate
possible extensions of the present work and promising
outlooks.
%
%
\section{\label{S2}Method}
In this section, we briefly review the now well established 
non-relativistic microscopic framework which leads to the results
on lead isotopes presented in Sect.~\ref{S3}.
It provides a natural setup for a treatment
of i) the mean-field properties by means of standard HF+BCS
or HFB methods with effective Hamiltonians
and ii) the collective dynamics of any amplitude by means of the GCM.
%
%
\subsection{ \label{S2.1} The Effective Hamiltonian and the Mean-Field Equations}
Over the last quarter of century, it has been acknowledged that the separation 
of the nuclear binding energy at the mean-field level in Hartree-Fock 
and pairing components in the spirit of the energy density
approach provides a fruitful starting point for an investigation
of the static and collective properties of nuclei.
Without requiring completely consistency, since the Hartree-Fock and
pairing fields remain coupled through the one-body
density matrix, this approach deviates
from the strict HFB theory as it can be performed with
the Gogny \cite{Gog75} or SkP \cite{DFT84} 
interactions. Still, in nuclear physics, 
the energy density approach is often
formulated in a way which preserves contact with the notion 
of effective force. Then, as it will be the case here, two
interactions are used; one in the particle-hole (Hartree-Fock) 
channel and one in the particle-particle (pairing) channel.
The introduction of these two effective forces
yields additional flexibility in the
phenomenological description of nuclear properties.  

Historically, the structure of the particle-hole interaction, 
which in heavy nuclei accounts for more than 99$\%$ of
the binding energy, has been investigated first. In the
present work, we use one of the latest versions of the
well established line of Skyrme interactions: the 
SLy4 parametrization \cite{CBH98} designed to describe properties
of nuclei (light and heavy), near the stability line or
in the neutron rich region up to the drip line with a similar accuracy.
Here, we are concerned with
stable and long lived lead isotopes. The quality of SLy4 for
such non exotic nuclei has been tested on numerous
examples \cite{CBH98,RBH99,FHS99,CNH99}. 
The force which acts in the pairing channel 
and defines the corresponding terms in the energy density
is taken from Refs.\ \cite{THB95,RBH99}. It is a zero-range 
interaction with a dependence on the nucleon space
density. This dependence ensures that the particle-particle two-body 
matrix elements are sensitive to the behavior of the single-particle 
orbitals mostly located at the surface of the nucleus.
This interaction has been tested on physical issues
sensitive to pairing properties, such as
the evolution of moment of inertia in extended
rotational bands as have been observed for instance in
superdeformed  \cite{THB95} and superheavy \cite{DBH01} nuclei.

Along a series of isotopes or isotones,
as either $N$ or $Z$ approaches a magic value, 
the mean-field method predicts a sudden
collapse of pairing. It is now
established that the magnitude of this transition
overemphasizes the lowering of pairing correlations
which must be expected when the single 
particle density at the Fermi surface
decreases significantly. 
Indeed, even in magic nuclei when the mean
value of the order parameter vanishes, it is expected 
that fluctuations of this order parameter
(in phase and magnitude) should still influence
nuclear properties. The importance of this so-called
dynamical pairing (as opposed to static pairing
in situations when the HFB and HF solutions differ
from each other) is one of
the questions that we investigate below within
the GCM formalism. Lipkin and
Nogami have proposed a prescription which 
attempts to correct the mean-field for the
effects of the phase fluctuations of the order
parameter \cite{Lip60,NZu64,PNL73}. This still not well
understood prescription has nevertheless
been shown to lead to global improvement (over
the mean-field theory) for the description of 
nuclear properties related to pairing. 
We will therefore also present results
obtained with the Lipkin Nogami prescription
which is also sometimes presented as an approximation
to the variation after projection method (VAP).

To conclude this section let us mention a technical point:
when pairing correlations are present, whether at the
static level (HFB) or because the Lipkin Nogami
prescription has been implemented, we have taken care that
no orbital of the continuum is occupied in
the HFB wave function. In practice, this is
achieved by means of a cutoff which excludes  all 
orbitals whose single particle energy
is 5MeV above the Fermi energy,
from the active pairing space.
%
%
\subsection{\label{S2.2}The Construction of the Collective Spaces}
In this work, the set of collective
wave functions $\{\vert\Phi(q)\rangle\}$
are determined by self consistent methods
($q$ denotes the value of the collective variable of interest).
For the modes associated with isoscalar
shape vibrations (monopole, qua\-drupole, octupole), these
spaces are obtained by constrained HF+BCS calculations with the appropriate
one-body operator as listed in Table 1 and $q$ is the expectation value of the
multipole moment operator
$q=\langle\Phi(q)\vert Q_{jm}\vert\Phi(q)\rangle$
\bt
\bc
\btab{ll}
\hline\noalign{\smallskip}
Vibration& Operator\\
\noalign{\smallskip}\hline\noalign{\smallskip}
Monopole   & $Q_{00}\propto
\sum_i r_i^2$\\
Quadrupole & $Q_{2m}\propto
\sum_i r_i^2\, Y_{2m}(\theta_i,\varphi_i)\q,m=0,\pm2$\\
Octupole   & $Q_{30}\propto
\sum_i r_i^2\, Y_{30}(\theta_i,\varphi_i)$ \\
\noalign{\smallskip}\hline\noalign{\smallskip}
\etab
\caption{\label{T1}
Constraining operators used to construct the collective spaces
associated with the lowest isoscalar nuclear shape vibrations. The 
quantities $r_i$, $\theta_i$ and
$\varphi_i$ refer to the spherical coordinates of 
the nucleon $i$ in an intrinsic reference frame whose
origin is at the 
nucleus center of mass. The $Y_{lm}$ are the 
standard spherical harmonics.
}
\ec
\et

For the vibrations in the pairing space
the collective coordinate must be
related to some global 
(complex) pairing gap. Since, as is discussed
in Sect.\ref{S2.3}, the fluctuations associated with the phase 
of the gap are 
taken into account by an exact projection after
the HF+BCS variation, there only remains to describe
the fluctuations of the gap magnitude. To build the 
associated collective space, there is no unique prescription
For instance, one may consider generating the space by means of
HFB calculations with a constraint on the fluctuations
of the neutron and proton numbers 
(i.e.\ the operators $\Delta\NH_\mu^2\,,\,\mu=n,p$) \cite{Egi00}.

In order to escape the technical difficulties associated
with the two-body nature of this operator, we have
adopted a second method. We construct the space by means of 
non constrained HF+BCS calculations using 
the same functional (with the SLy4 parametrization)
for the mean-field part and replacing the pairing
functional by that associated with 
an auxiliary Hamiltonian depending on
two real gaps parameters $\Delta_n$ and $\Delta_p$: 
\begin{equation}
{\hat H}_P(\Delta_n,\,\Delta_p)=
\frac{\Delta_n}{2}\sum_{\mu=1}^N\, a^\dagger_{n\mu}a^\dagger_{n{\bar\mu}}
+
\frac{\Delta_p}{2}\sum_{\mu=1}^Z\, a^\dagger_{p\mu}a^\dagger_{p{\bar\mu}}
+\ \rm{c.c.}\quad,
\end{equation}
where $a^\dagger_{\tau\mu}$ is the creation operator of the nucleon
of isospin $\tau$ in the individual state $\mu$ and where $\bar\mu$ denotes
the time reversed orbital. The variation of the corresponding total functionals
generates a set of BCS states $\{\vert\Phi(\Delta_n,\Delta_p)\rangle\}$.
Once this collective space has been determined, 
all further calculations, (expectation values
and GCM matrix element calculations (see Sect.~\ref{S2.3}))
are performed with the Hamiltonian described in Sect.~\ref{S2.1}.
This method has already been used in Ref.\ \cite{MBD91} to investigate
the influence of dynamical pairing on
tunneling probabilities between the superdeformed and normal
wells in $^{192}$Hg. 

The GCM method involves a non orthonormal collective basis.
The two prescriptions for constructing the pairing
collective space sketched above give should therefore lead to similar
results (collective energy and wave functions) once the
bases are large enough. In addition, the GCM
method does not explicitly depend on which collective
coordinates have been selected to span the total collective
space. As long as the mapping is one to one, 
rather than by $(\Delta_n,\Delta_p)$, we may
as well choose to plot the GCM collective properties in
terms of the coordinates
$(\langle\Delta\NH_n^2\rangle,\langle\Delta\NH_p^2\rangle)$
where $\langle\AH\rangle$ denotes the expectation
value of the operator $\AH$ in the state 
$\vert\Phi(\Delta_n,\Delta_p)\rangle$ (or by any
other set of coordinates).
%
%
\subsection{\label{S2.3}The Projection and the Collective Dynamics}
The Lipkin-Nogami prescription determines a BCS wave function
which, in priciple, takes into account the pairing
fluctations associated with the phase $\phi$ of the pairing.
This wave function can be used directly to evaluate
the expectation value of any operator as proposed in the
original articles of Lipkin and Nogami. In this work,
such values will be refered to as L.N.\
One can also consider that this BCS state is
an approximation of the exact VAP state.
In order to test the validity of this assumption,
one must then extract from the LN-BCS state its
component with correct (i.e.\ $N_0$) particle number
by means of a projection $\PNz$. 
Henceforth, the results associated
with such a two-step approach will be labeled L.N.\ (proj.) 

The projection
\begin{equation}\label{proj}
\PNz
= \usp\int_{-\pisd}^{\pisd} {\rm d} \phi \,\eee{i\phi(\NH-\Nz)}\q,
\end{equation}
involves an integral over $\phi$ with the weight factor $\eee{-i\phi\Nz}$. 
From the mean-field wave functions \{ $\kphiq$ \} associated with
the value $q$ of the collective variable, we build the states    
\begin{equation}
\label{projwf}
\kphin
= \PNz\kphiq\q,
\end{equation}
which  form the non orthogonal projected collective basis \{$\kphin$\}
The expectation value $\EC(q)$ of the Hamiltonian in this basis is 
the projected deformation energy curve:
\begin{equation}
\label{projeq}
\EC(q)
= \frac{\bphind\vert\HH\kphin}{\bphind\kphin}
= \frac{\bphiq\vert\HH\PNz\kphiq}{\bphiq\vert\PNz\kphiq}\ .
\end{equation}
Several such curves are discussed in Sect.\ref{S3}.
The value at the minima of $\EC(q)$ yields an
energy corresponding to a restricted VAP
(a variation limited to the subspace 
spanned by the collective variable $q$). 
The quality of this upper bound to the VAP energy depends on the
relevance of the collective space for dynamical 
pairing correlations.

In a next step, we consider a more general N-body wave function
defined as a linear superposition of projected HFB states with 
an unknown weight function $f_j(q)$: 
\begin{equation}
\label{gcmwf}
\kpsi
= \int \! {\rm d}q \, f_j(q) \, \kphin \q.
\end{equation}
In this definition, the label $j$ recalls that
several states $\kpsi$ are obtained corresponding to
the correlated ground state and 
to the collective excited states.
Within the GCM method, the function $f_j$
is determined by a variation
of the total energy $E_j$
\begin{equation}
\label{gcme}
E_j
= \frac{\bpsi\vert\HH\kpsi}{\bpsi\kpsi}\q,
\end{equation}
with respect to the function $f_j^*(q)$. This leads
to the Hill-Wheeler equations \cite{HWh53}. When particle
number projection is imposed on each basis state, the
kernels $\UC$, $\NC$ and $\HC$ of the integral operators
entering this equation are given by
\begin{equation}\label{gcmk}
{\left(\ba{c}
\UC(q',q)\\
\NC(q',q)\\
\HC(q',q)
\ea\right)}=
\usp\int_{-\pisd}^{\pisd}\,{\rm d} \phi\,
\eee{-i\phi\Nz}\langle\Phi(q')\vert
{\left(\ba{c}
\uni\\
\NH\\
\HH
\ea\right)}
\eee{i\phi\NH}\kphiq
\end{equation}
The evaluation of these $(q',q)$ dependent
kernels is the most time consuming numerical part
(in fact, there are four of them
since neutron and proton numbers have
to be conserved separately).
They involve a double integral (one for $N$
and one for $Z$) of matrix elements of one-body and
two-body operators between all possible states of the collective
basis $\{\kphiq\}$. The formulae useful for the
calculation of the sets of matrix elements are
given in Ref. \cite{BDF90}. 
Other kernels involving multipole
moment operators associated with the various deformations
are also needed for the computation of the shape
properties of the collective GCM states.
The reference just quoted
also recalls how the solution of the Hill-Wheeler
equation can be reduced to the diagonalization of
an hermitian matrix computed from the kernels
$\UC$, $\NC$ and $\HC$.

The functions $f_j$'s do not form an orthogonal set.
Using the overlap kernel, one can define
the functions $g_j$'s
\begin{equation}\label{collwf}
g_j(q)=(\UC^{1/2}.f_j)(q)\q,
\end{equation}
which are orthonormal and can be interpreted as
collective functions in the usual sense. 
In Eq.~(\ref{collwf}), the notation $\UC^{1/2}$
stands for the integral operator whose square is equal to
the integral operator with kernel $\UC(q',q)$.
In the next section, we present some functions 
$g_j$ associated with the
quadrupole and octupole vibrations.

In the case of quadrupole vibrations, the GCM is solved
separately in each representation of the permutation
group of the three intrinsic axes. As discussed in Ref.\ 
\cite{BDF91,HBDF93}, the two independent representations
of this group have different angular-momentum-parity ($J^\pi$)
contents: in the fully symmetric one, the major component
of the collective wave function corresponds to $0^+$
spin and parity, while, in the
two dimensional representation, the collective N-body
wave function is predominantly of the $2^+$ type. Very
recently, we have contructed an exact angular momentum projection code 
\cite{VHB00} and we have quantitatively checked the quality of this
approximate spin projection. We have found that it is accurate
whenever the extension of the collective function $g$
does not extend beyond the value \mbox{$\langle Q_{20}\rangle=4$}b
of the quadrupole deformation. Since the influence of
the spherical magic proton number \mbox{$Z=82$},
insures that $g$ is non zero only in
the close vicinity of the $\langle Q_{20}\rangle=0$ point, this is
the case for the $g$ functions of the lead isotopes.

Although our HFB and GCM codes 
allow triaxial deformations, the results presented below
are based on collective bases including only
prolate and oblate states. This is justified by  the
small extension of the collective wave functions.
Nonetheless, all the GCM calculations 
presented below mix states located
on the six semi-axes corresponding to the triaxiality 
angles $\gamma=0$, $\pm 60^\circ$, $\pm 120^\circ$ and $180^\circ$. 
We have tested the quality of this approximation by a performing
full triaxial calculations on a few cases and we have 
checked that our conclusions do not depend on this degree of freedom.

The mean-field (and Lipkin-Nogami) wave functions spanning the collective
spaces are constructed from an energy density functional
whose particle-hole and particle-particle components are defined by
means of two distinct effective density dependent two-body
forces. For the sake of simplicity, we have presented the 
standard GCM formalism with a single effective 
Hamiltonian. In practice, we have adapted the GCM
to the  choice made in Sect.~\ref{S2.1} and defined 
a generalized energy density to be used 
in the computation of the energy kernel $\HC(q',q)$ given by
Eq.~(\ref{gcmk}). This involves both a prescription for handling
the density dependence of the force and an extension of
the energy density formalism to the non-diagonal matrix
elements which appear in the integral in Eq.~(\ref{gcmk}).
We use the method described in Ref.\ \cite{BDF91} (for an 
alternative choice, see Ref.\ \cite{ERo92} for instance).
The formulae for the two terms
of the extended energy density are given in Ref.\ \cite{BDF91}
and the implications and potential problems 
associated with this formulation are discussed
in Refs.\ \cite{TFB92,Egi00}.
%
%
\section{\label{S3}Results}
We consider successively, the influence of correlations on
the ground state properties and the collective excitations
of five lead isotopes.
%
%
\subsection{\label{S3.1} Ground State Correlation Energies}
\begin{figure}[t!]\begin{center}
\includegraphics*[scale=0.4,angle=-90.,draft=false]{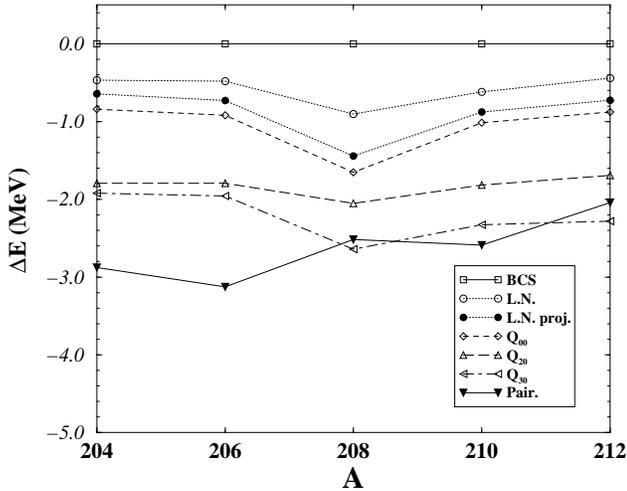}
\caption{\label{BFHVfig1}
Differences between the calculated 
total binding energy of the ground state of 
the five lead isotopes and a BCS reference. 
The labels $Q_{i0},\, i=0,2,3$ refer to GCM calculations
taking into account the different types of shape
collective excitations reported in Table~\ref{T1}, while
``Pair.'' stands for the pairing vibrations.
Except for the curves BCS and L.N., all
results involve an exact projection on $N$ and $Z$.
}\end{center}\end{figure}

In Fig.~\ref{BFHVfig1}, for each of the isotopes
$^{204}$Pb, $^{206}$Pb, $^{208}$Pb, $^{210}$Pb and $^{212}$Pb, 
we have plotted the difference
between the binding energies of the ground state calculated
by GCM on several collective spaces and a BCS reference value. 
In practice, for the five nuclei,
the proton part of this reference wave function reduces
to a Slater (HF) determinant. This is also true
for the neutron component in $^{208}$Pb. 
In addition, in Fig.~\ref{BFHVfig1}, we have plotted the difference
between the same reference and the binding energies calculated
by means of the Lipkin Nogami method with and without a
projection on the neutron and proton numbers.

The ordinate scale of the figure shows that, irrespective of the nature
of the collective vibration, the correction to the total binding energy 
is small (about 3 MeV). Considering the effective character of the two-body 
interactions defining the energy density, 
this is a  satisfactory result. As we discussed in the introduction,
one can consider either that, to first order
effects of correlations can be effectively  included
into the parametrization of the force, or conversely 
that the presently available
forces have to be readjusted so as to underbind slightly with
respect to data. This would leave room for the small corrections 
associated with correlations
to be computed on top of the mean-field solution,
as we have done. In Ref.\ \cite{RFr85}, much larger values of correlation
energies have been found. However, as was pointed out in this
reference, the method 
based on a sum rule approximation
to the RPA suffers from an ambiguity related to 
the double counting in RPA formulas
for the correlation energies. The GCM
method is free from such a problem.

The relative magnitudes of the correlations tell about the 
nuclear sensitivity to various collective modes.
For lead isotopes the octupole mode is the most
efficient among the three nuclear shape vibrations analyzed
in this work. One also notes that even for nuclei close to
$N$ and $Z$ magic shell closures, pairing vibrations
generate the largest correction.

The L.N.\ and even more the
the L.N.\ projected curves show a larger effect for the 
doubly magic nucleus $^{208}$Pb. For the latter curve,
the difference with BCS increases from about
0.65 MeV at $^{204}$Pb and $^{212}$Pb to 1.4 MeV at
$^{208}$Pb. A possible interpretation is that the
LN prescription which is designed to correct for vanishing
pairing fluctuations in the HF+BCS solution,
is more effective in $^{208}$Pb
when the single particle energy gap at the Fermi surface is
larger. With respect to the
second curve (L.N.), the monopole,
and octupole vibrations introduce an additional lowering
of the energy whith an almost $N$ independent value equal
to 0.2 MeV and 1.4 MeV respectively. By contrast, the
curve associated with pairing vibrations shows an
overall decrease versus $N$ of 
the binding energy correlations : over five isotopes it
is reduced by about 0.6 MeV.

\begin{figure}[b!]\begin{center}
\includegraphics*[scale=0.4,angle=-90.,draft=false]{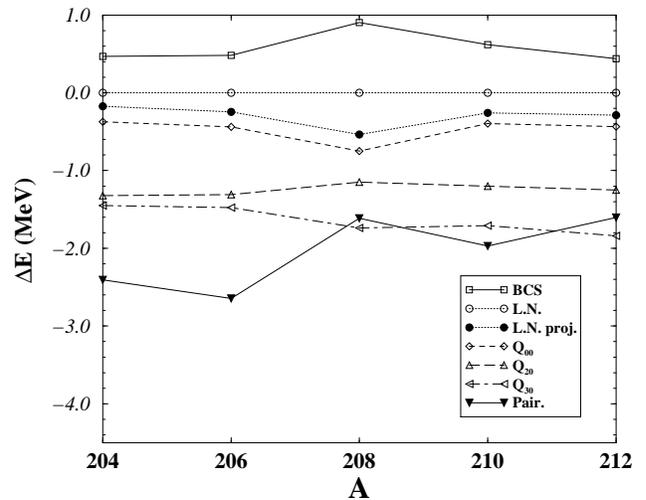}
\caption{\label{BFHVfig2}
Differences between the calculated 
total binding energy of the ground state of 
the five lead isotopes and the
Lipkin Nogami reference. 
The labels $Q_{i0},\, i=0,2,3$ refer to GCM calculations
taking into account various types of shape
collective excitation as described in Table~\ref{T1}, while
``Pair.'' stands for the pairing vibrations.
Except for the curves BCS and L.N., all
results involve an exact projection on $N$ and $Z$.
}\end{center}\end{figure}

In Fig.~\ref{BFHVfig2}, we show the same results
when the standard LN energy is taken as
a reference. This presentation 
illustrates the improvements brought about by
an exact projection as compared to the
approximate prescription introduced by Lipkin and Nogami.

When the LN prescription
does not yield a good approximation of
the binding energy of the particle projected VAP solution, one
could hope it to be a practical method to construct a BCS state close to
the VAP intrinsic state. Then, a projection of this BCS
wave function should give almost the same energy as a VAP calculation.
Our results show that the validity of this property of the LN prescription
is at least nucleus dependent.
The minimum of the energy surface obtained
with the projected BCS states spanning the pairing
collective space  provides
an upper bound of the VAP energy. From
the results given in Table \ref{T2}, it is seen that 
for the lighter lead isotopes, it is more than 1MeV below the
energy associated with the projected Lipkin Nogami state.
The same table shows that the lowest binding energies
correspond to projected BCS states with particle number
fluctuations larger than those predicted by the LN prescription.
For $^{208}$Pb and isotopes above, the differences between the projected 
LN and approximate VAP solutions are smaller.

\bt\bc\btab{cccc}
\hline\noalign{\smallskip}
$A$&Lipkin-Nogami        & Variat. Minimum      & $\Delta E$\\
   &(\,$\DNn$,\,$\DNz$\,)&(\,$\DNn$,\,$\DNz$\,)& (MeV)\\
\noalign{\smallskip}\hline\noalign{\smallskip}
204&(\,2.54,\,1.14\,)    &(\,2.90,\,1.61\,)    & $-1.51$\\
206&(\,1.85,\,1.17\,)    &(\,2.60,\,1.95\,)    & $-1.36$\\
208&(\,1.36,\,1.17\,)    &(\,1.92,\,1.12\,)    & $-0.21$\\
210&(\,1.92,\,1.17\,)    &(\,2.37,\,1.13\,)    & $-0.84$\\
212&(\,2.59,\,1.17\,)    &(\,2.58,\,1.50\,)    & $-0.28$\\
\noalign{\smallskip}\hline\noalign{\smallskip}
\etab
\caption{\label{T2}
Neutron and Proton particle number fluctuations of the
BCS states associated with the Lipkin Nogami solution and
the minimum of the variational space constructed for the
study of pairing vibrations (see text). The quantity
$\Delta E$ give the difference between the 
energy of the latter state and the LN state 
{\it after particle number projection has been effected}
(i.e.\ L.N.\ proj.). 
}
\ec\et

\begin{figure}[b!]\begin{center}
\includegraphics*[scale=0.4,angle=-90.,draft=false]{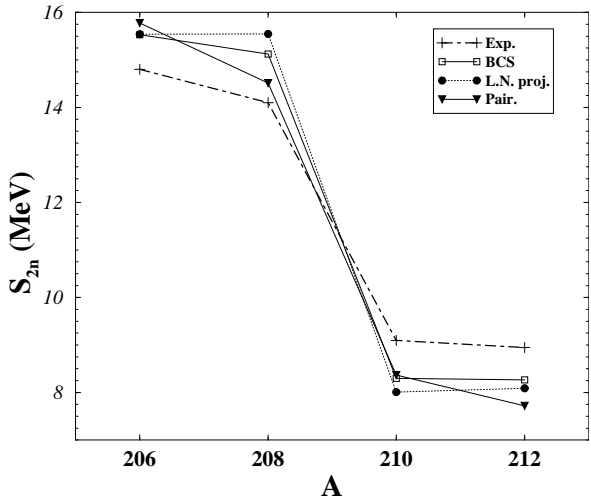}
\caption{\label{BFHVfig3}
Comparison between the experimental 
two neutron separation energies of the lead isotopes 
and the values calculated within the BCS, projected LN
and GCM methods. In the latter case, the collective
space allows fluctuations
of the gap parameter.
}\end{center}\end{figure}

The discontinuity of the two-neutron
separation energy $S_{2n}$ at a shell closure
predicted by  mean-field methods 
is often  larger than found experimetally.
This effect can be seen in Fig.~\ref{BFHVfig3}.
The magnitude of the drop of the BCS $S_{2n}$ value between $^{208}$Pb
and $^{210}$Pb exceeds data by approximately 1.8 MeV. Although 
not shown in the figure, we have checked that
the differences between the BCS and experimental
$S_{2n}$'s are not larger than 0.2 MeV for mass number $A$
smaller than $204$. Therefore the larger discrepancy at 
$A=208$ cannot be ascribed to the asymmetry of the effective force 
SLy4 and suggests a deficiency of the HF+BCS method
for magic or near magic nuclei. 
This assumption is confirmed by the fact that 
the $S_{2n}$'s calculated with the GCM on a
collective space allowing the fluctuations of
the pairing gap leads to an improvement by about
0.6 MeV (i.e. 33$\%$). By contrast, the Lipkin-Nogami prescription
does not improve the mean-field results although it is supposed to 
approximately take into account the effect of these pairing vibrations.
We have not plotted the curves with the energies
resulting from the GCM associated with
shape vibrations. Indeed,
they are not much different from the LN curve.
%
%
\subsection{\label{S3.2} Collective Functions and Excitation Spectrum}
\begin{figure}[t!]\begin{center}
\includegraphics*[scale=0.4,angle=-90.,draft=false]{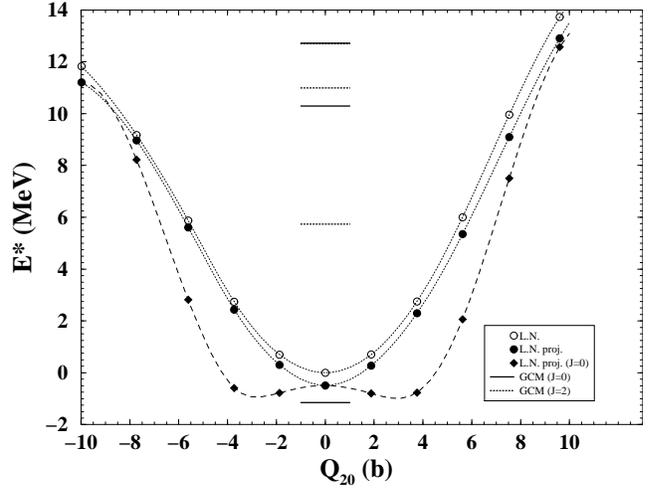}
\caption{\label{BFHVfig4}
Quadrupole deformation energy curves of $^{208}$Pb calculated 
with the Lipkin-Nogami state without or with exact projection 
on the neutron and proton particle number. In the 
latter case, the curve in the subspace containing 
the $0^+$ component of the 
deformed function is also given. The solid straight segments give 
the GCM collective energy in this latter subspace, while 
the dotted segments give the energy in the subspace containing
the $2^+$ components of GCM the collective functions.
}\end{center}
\end{figure}
The $Z=82$ and $N=126$ magic numbers have a definite influence
on the individual and collective properties of the five 
isotopes investigated in this work. They 
strongly favor a mean-field solution corresponding to
a HF (no pairing) spherical wave function. 
The Lipkin Nogami prescription
which transforms the HF states into HF+BCS ones by
enforcing a smooth decrease of 
the occupation probabilities
from the hole to the particle states 
does not modify substantially the deformation properties
of the collective energy surfaces. 
As it can be seen in Fig.~\ref{BFHVfig4},
the LN energy curves of $^{208}$Pb,
with and without exact particle
number projection, exhibit a
well marked minimum at the spherical point.
The curves are very similar, the
main effect of the projection being a downward shift by about
0.5 MeV. When an approximate angular projection on the
$0^+$ state is performed, the bottom of the
well is flattened over an extension of $\pm 4$b.
Qualitatively, such an effect 
is expected on general grounds. The 
LN wave function at the spherical point is a pure
$0^+$ state and is unaffected by projection while
a deformed mean-field wave function is a mixture of different
spin components. The energy of 
the $0^+$ component is lower than 
that of all other spins and therefore than
that of the deformed HF+BCS wave function.

The GCM energies of the lowest collective
quadrupole (approximate) $0^+$ and $2^+$ states of $^{208}$Pb
are also plotted in Fig.~\ref{BFHVfig4}. 
The first excited state is a $2^+$.
A comparison with the GCM monopole 
calculation shows that the lowest $0^+$ collective state in
the spectrum of $^{208}$Pb is  the one shown 
in Fig.~\ref{BFHVfig4}, the
energy of the first excited monopole GCM state being at 13.4MeV.
This latter value which does not vary with $N$ over
the five isotope series, agrees 
with monopole giant resonance data \cite{Bra87,YCL99}.
This nice result is related to the nuclear matter 
incompressibility of the SLy4 force, 220 MeV, 
a value which is known
to be coherent with experiment \cite{Bla80,Spe91,BBD95}.

\begin{figure}[b!]\begin{center} 
\includegraphics[bb=92 50 570 672,
scale=0.4,angle=-90.,draft=false]{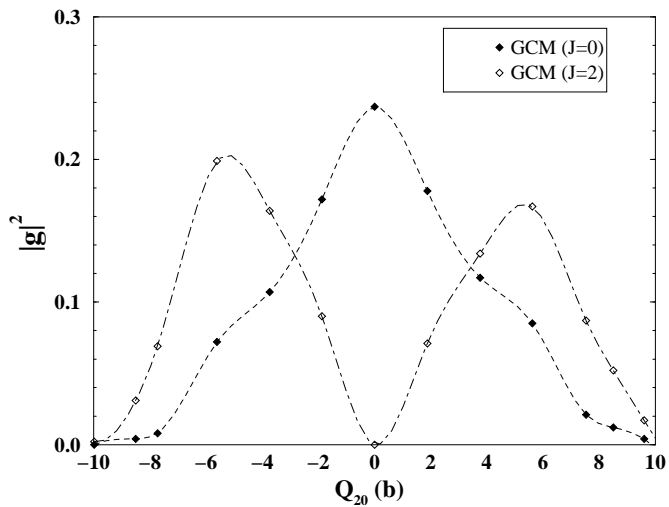}
\caption{\label{BFHVfig5}
GCM collective functions associated with 
quadrupole energy deformation of $^{208}$Pb. 
The solid curve gives 
the wave function of the lowest energy state
in the subspace containing the $0^+$
component of the total wave function, while 
the dashed curve gives the same information for
the subspace with the $2^+$ components.
}\end{center}\end{figure}
Fig.~\ref{BFHVfig5} displays the GCM probabilities (i.e.\
$\vert g\vert^2$ where $g(q)$ is 
defined by Eq.~(\ref{collwf})) for the first $0^+$ and
$2^+$ states. The former is peaked at the spherical
point. Its width ($\approx 4$b) corresponds to the quadrupole
extension of the bottom of the angular momentum
projected energy curve shown in Fig.~\ref{BFHVfig4}.
The collective probability distribution  of the
first $2^+$ GCM state displays a node at $Q=0$. Its extension
is almost symmetric on the prolate and oblate sides
with a slight preference for the oblate deformation. 

\begin{figure}[t!]\begin{center}
\includegraphics*[scale=0.4,angle=-90.,draft=false]{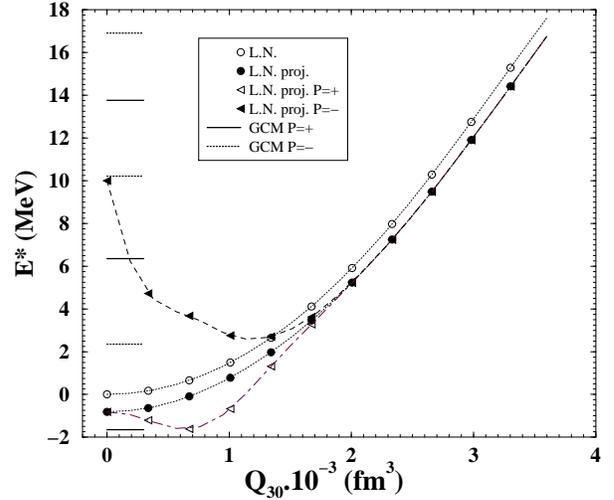}
\caption{\label{BFHVfig6}
Octupole energy deformation energy curves calculated 
with the Lipkin-Nogami state, without or with exact projection 
on the particle number. In the latter case, the curve in 
the subspace containing the positive 
and negative parity components of the 
deformed function is also given. 
The solid (resp. dotted) segments give 
the GCM collective energy in the 
positive (resp. negative) parity
subspace of the GCM collective functions.
}\end{center}\end{figure}

Four octupole deformation related energy curves for $^{208}$Pb
are plotted in Fig.~\ref{BFHVfig6} 
against positive expectation values
of the constraining operator $Q_{30}$ (see Table~\ref{T1}).
Since this operator is parity negative, 
the negative abscissae part of each curve
can be deduced by symmetry with respect to the ordinate
axis. The LN and LN proj.\ curves 
display a minimum at the spherical
point $\langle Q_{30}\rangle=0$. The projection merely
increases the binding energy by 0.6MeV. An additional 
projection of the wave function on the 
total positive (resp. negative) parity separates
the subspace which contains the $0^+$ (resp. the $3^-$)
component of the LN particle projected state. The LN and LN proj. 
curves only differ significantly
for small values of $\langle Q_{30}\rangle$ ($<2.10^3$fm$^3$), 
namely as long as the overlap between 
the LN particle-projected wave function
and its parity reversed is non zero.
The positive parity curve has a minimum for 
$\langle Q_{30}\rangle\approx 10^3$fm$^3$. Below,
we will see that this number 
provides a measure of the octupole fluctuations
in the $0^+$ GCM state. We recall that
symmetry garantees that the average value of $\langle Q_{30}\rangle$ 
vanishes exactly for all GCM states and, in addition, that
the negative parity projected 
energy curve diverges at $\langle Q_{30}\rangle=0$.

\begin{figure}[t!]\begin{center}
\includegraphics[bb= 92 50 575 672,
scale=0.4,angle=-90.,draft=false]{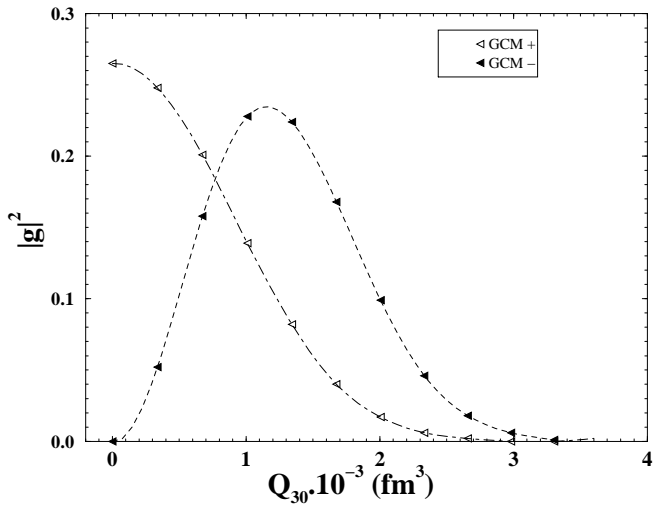}
\caption{\label{BFHVfig7}
Collective wave function of the lowest energy state with a
given parity associated with 
the octupole deformation of $^{208}$Pb . 
The solid (resp. dashed) curve gives 
the ground state function in the subspace containing the
positive (resp. negative) parity component of the 
wave function.
}\end{center}\end{figure}

The GCM energies given in Fig.~\ref{BFHVfig6} 
display an alternating sequence of positive and negative parity
states. The first octupole GCM
excited state should correspond to the
lowest experimental $3^-$. Its probability distribution
($\vert g(q)\vert^2$) and that of
the $0^+$ are plotted in Fig.~\ref{BFHVfig7}.
They give an information on the magnitude
of the fluctuations in the corresponding collective state.
Returning to Fig.~\ref{BFHVfig6}, one checks 
that each of these two wave functions
displays the features expected for the
quantum ground state in the projected
Lipkin Nogami curve with the associated parity.
 
Up to now, the discussion in this subsection 
which centered on deformation energy curves and
collective probability distributions has
been concerned only with the doubly magic
nucleus $^{208}$Pb. The same curves 
calculated for the four other isotopes
are qualitatively  similar while, they yield significantly 
different collective energy spectra.Although one can discern 
evolutions of the energy curves versus $N$ which are at the origin 
of the differences found in the GCM spectra, they 
turn out to be neither spectacular nor linked to any
specific part of the collective space. For this reason,
we chose not to plot them; they only
display a smooth modification 
of the same global pattern along the isotope series.

\begin{figure}[t!]\begin{center}
\includegraphics*[scale=0.4,angle=-90.,draft=false]{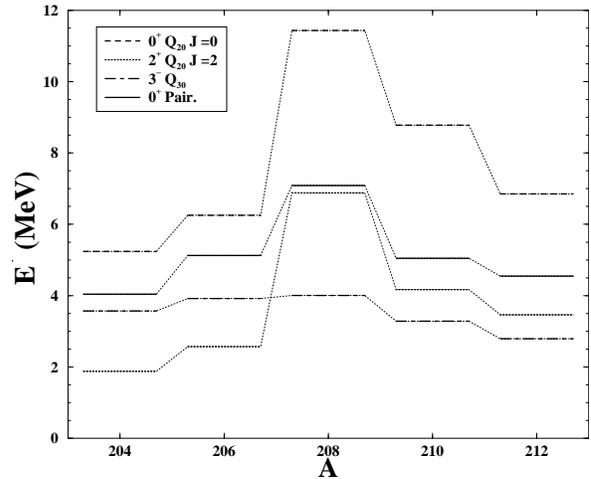}
\caption{\label{BFHVfig8}
Evolution of the lowest excited collective states of five lead
isotopes near the magic nucles $^{208}$Pb. The figure compares data
with results of GCM including an exact projection on 
particle ($N$ and $Z$) numbers and an aproximate projection
on angular momentum.
}\end{center}\end{figure}

In Fig.~\ref{BFHVfig8}, the collective GCM spectra for the
three first excited states ($0^+$ and $2^+$ quadrupole,
$3^-$ octupole) are given. Qualitatively, the overall pattern 
of data is well reproduced. The positive parity levels are pushed 
to high energy as one crosses \mbox{$N=126$}. As a consequence,
the lowest collective state in $^{208}$Pb is a $3^-$. As one
moves away from the shell closure, the collective spectrum
returns to the usual midshell pattern with the
$2^+$ state being the lowest. However,
the collective states whether quadrupole ($0^+$
and $2^+$) or octupole are found by the GCM at 
energies between 1.5 to 2 larger than observed. 
Although the formalism of the GCM does not distinguish
potential from collective inertia effects, if we were to adopt
the concepts of the standard collective model
to analyze our discrepancy with data, we could say that
the results are consistent with the calculated moments 
of inertia of the low energy shape vibrations being twice too small.
In the next section, based on the experience derived from
Refs.\ \cite{PZP99,ZPP99} and from the existence of
significant dynamical pairing effects evidenced
by the results discussed in Sect.~\ref{S3.1}, we put forward
possible explanation for the GCM results.

With the Skyrme interaction SLy4,
the quadrupole collective transition matrix element $^{208}$Pb
$B(E_2)\uparrow$=7.1 W.u.\ is close to the data
(8 W.u.). The agreement is as
good for the octupole transition since we find
$B(E_3)\uparrow$=18.7 W.u.\ as compared to the
experimental value equal to 32 W.u.\ \cite{Bra87,Dja80,Mar86,YCL99}.
We note however that these results are very sensitive to
the interaction since a similar calculation with the
SGII interaction yields 14 W.u.\ and 58 W.u.\ respectively.

%
%
\section{\label{S4}Conclusions and Perspectives}
We have performed an analysis of the collective correlations
and modes of excitations in the vicinity of the magic nucleus 
$^{208}$Pb. One of the characteristics of this work is an
attempt to maximize the coherence of the description of
the mean-field and collective effects for any amplitude.
This is done by means of the GCM solved in collective bases.
Depending on the type of collectivity, the GCM bases are obtained
from constrained calculations (shape vibration) or derived
from a set of auxiliary Hamiltonians (pairing vibrations). Another
element of consistency is the use of the same effective 
particle-hole (p-h) and particle-particle (p-p) interactions
in the complete sequence of calculations. These two forces
had been determined in earlier analyses of ground state and 
high spin properties by means of mean-field methods.

We find that correlation effects on the total binding 
energies are small. A posteriori, this result justifies 
the perturbative-like calculation of such correlations
on top of the mean-field solution as done in the
present work. It also supports attempts at
the definition of a force to be used
exclusively in mean-field calculations and  designed  to 
effectively incorporate the small effect of 
correlations \cite{Ton00,Bro98}. The latter type of force is
a practical tool for the extensive nuclear mass tabulations 
required for a global analysis of astrophysical nucleosynthesis. 

Furthermore, we have found that pairing vibrations induce a softening
of the rapid variation of the
$S_{2N}$ curves at the crossing of the magic shell \mbox{$N=126$}.
Such accidents of the nuclear mass chart have a significant
influence on the understanding of nuclear abundance
peaks resulting from high neutron fluxes in stellar events such as
class II supernovae explosions. A check of the persistence
of this effect at the points where the r-process path crosses
the \mbox{$Z=82$} magic line (on the $N$-rich side) appears 
therefore desirable. 

Our work also provides an illustration of the validity of
the Lipkin Nogami prescription. 
The LN method should give good approximations of both the
VAP wave function and energy. More restrictively, it is 
considered that even when the LN energy
does not reproduce well the VAP energy, 
an exact projection performed on the LN wave function
does so \cite{MCD93}. This implies that VAP
is well approximated by a projection after the solution of
the LN equations. Our calculations show the limitations
of these two assumptions. We find that 
even after projection the discrepancy on the binding energy
is nucleus dependent and can be larger than 1.0MeV. Our calculation
provides only a lower bound of this discrepancy since the variational 
space used in the present work only covers a fraction of the space
available to a full VAP calculation

We find that, for the five lead isotopes,
the GCM collective vibrations associated with
our variational spaces only generate small fluctuations around
the spherical shape and do not modify significantly the 
expectation values of multipole moments. The behaviour of the
low energy collective spectrum is qualitatively correct.
In particular the relative positions 
of the $0^+$, $2^+$ and $3^-$ states along the isotope sequence is
well reproduced. We also find a reasonable agreement on
electromagnetic transition matrix elements.
However, the energy scale of our spectrum
is too large by a factor between 1.5 and 2 compared to data. 
The reason for this discrepancy is not yet clear. 
This is all the more disconcerting that RPA calculations
on the Hartree-Fock solution using the same interaction 
provide a much better agreement with the low energy collective 
spectrum of $^{208}$Pb \cite{Rei01}. For this nucleus, we have 
additionally performed a GCM calculation of quadrupole and 
octupole vibrations using the SGII interaction for which
RPA calculations are also available \cite{Col01}.
We find a similar pattern of fair agreement on transition matrix
elements and discrepancy on collective spectrum energies.

From these results, some directions for further work can be outlined. 
First it appears useful to extend the calculation to other regions 
of the mass table to test the importance of pairing
vibrations at shell closure. In particular
one should consider other proton magic numbers with
a focus on the neutron rich side at places of interest for
the r-process. Second, one sould  extend the formalism
with a coupling between the vibrational modes which have been detected
to be especially active. For the lead isotopes, 
the first such investigation should be devoted to a
coupled calculation of pairing and octupole vibrations.

An other point which needs clarification is the discrepancy
on the low energy collective spectrum density. The
problem does not seem related with the interaction since
RPA provides a rather good agreement with data. We note also that
with the same forces, GCM and RPA results on giant resonances
agree together (and also with data). A first reason for the
differences on low energy collectivity may be due 
to the use of a GCM basis of wave functions including dynamical pairing
while RPA is performed on a Hartree Fock solution.

A discussion of this problem can be found 
in  \cite{PZP99,ZPP99}. Indeed, in cranking
type calculations, pairing reduces collective mass
parameters and leads to a decompression of the spectrum.
In such a case, one would have to reconsider which
fraction of the pairing must be assigned to static (i.e.\ BCS
or HFB) or dynamic (fluctuations of the magnitude
of pairing gap) correlations. Another source for the different
GCM and RPA results may be due to the definition of the GCM collective 
spaces. To study shape vibrations, we have used a collective
basis generated by self consistent calculations with constraining 
operators proportional to multipole moments. A description
of the low energy mode could require other radial form factors
more peaked at the nuclear surface than the present $r^n$ form 
factors. The RPA transition form factors may provide indications
for more appropriate constraining operators. 

As a conclusion, the present work has demonstrated the interest of our 
approach and also shown  some of its limitations in its present form.
Nevertheless we believe that it establishes 
that the GCM performed on many-body bases including exact symmetries 
such as particle number, parity (as is done here) but also angular momentum
has strong potential for the description of 
nuclear properties which remains largely to be explored.
%
%
\subsection*{Acknowledgements}
We thank P.--G.\ Reinhard and G.~L.\ Colo for providing us information on
specific RPA calculations. This work has been supported in part
by the PAI--P3--043 of the Belgian Office for Scientific Policy.
%
%

\end{document}